\documentclass[conference]{IEEEtran}
\IEEEoverridecommandlockouts
\usepackage{cite}
\usepackage{amsmath,amssymb,amsfonts}
\usepackage{graphicx}
\usepackage{textcomp}
\usepackage{algorithm}
\usepackage[noend]{algpseudocode}
\usepackage{multirow}
\usepackage{adjustbox}
\usepackage{booktabs}
\usepackage[table,xcdraw]{xcolor}
\usepackage{makecell}

\usepackage{listings}
\definecolor{dkgreen}{rgb}{0,0.6,0}
\definecolor{gray}{rgb}{0.5,0.5,0.5}
\definecolor{mauve}{rgb}{0.58,0,0.82}

\usepackage{mathtools}
\DeclarePairedDelimiter{\ceil}{\lceil}{\rceil}

\def\BibTeX{{\rm B\kern-.05em{\sc i\kern-.025em b}\kern-.08em
    T\kern-.1667em\lower.7ex\hbox{E}\kern-.125emX}}
\begin{document}

\title{
ITERA-LLM: Boosting Sub-8-Bit Large Language Model Inference via Iterative Tensor Decomposition
\thanks{* Both authors contributed equally to this work.}
\vspace{-10mm}
}
\author{\IEEEauthorblockN{Yinting Huang*, Keran Zheng*, Zhewen Yu, Christos-Savvas Bouganis}
\IEEEauthorblockA{\textit{Department of Electrical and Electronics Engineering} \\
\textit{Imperial College London}\\
London, United Kingdom \\
\{justin.huang20, keran.zheng18, zhewen.yu18, christos-savvas.bouganis\}@imperial.ac.uk}
\vspace{-10mm}
}

\maketitle

\begin{abstract}
Recent advancements in Large Language Models (LLMs) have demonstrated impressive capabilities as their scale expands to billions of parameters. Deploying these large-scale models on resource-constrained platforms presents significant challenges, with post-training fixed-point quantization often used as a model compression technique. However, quantization-only methods typically lead to significant accuracy degradation in LLMs when precision falls below 8 bits. This paper addresses this challenge through a software-hardware co-design framework, \textit{ITERA-LLM}, which integrates sub-8-bit quantization with SVD-based iterative low-rank tensor decomposition for error compensation, leading to higher compression ratios and reduced computational complexity. The proposed approach is complemented by a hardware-aware Design Space Exploration (DSE) process that optimizes accuracy, latency, and resource utilization, tailoring the configuration to the specific requirements of the targeted LLM. Our results show that \textit{ITERA-LLM} achieves linear layer latency reduction of up to 41.1\%, compared to quantization-only baseline approach while maintaining similar model accuracy.
\end{abstract}


\section{Introduction}
\label{sec:section1}
The rapid advancement of Transformer-based Large Language Models (LLMs) has revolutionized a wide range of Natural Language Processing (NLP) tasks. A phenomenon observed in recent research is the emergent capabilities of LLMs as they scale to billions of parameters\cite{hoffmann2022trainingcomputeoptimallargelanguage}. However, supporting the unprecedented scale of LLMs introduces significant challenges, particularly in terms of computational and memory resources. 

While GPUs have been the primary focus for optimizing LLM inference, they are often constrained by high power consumption and poor performance for latency-sensitive workloads. Contrary to GPUs, Field-Programmable Gate Arrays (FPGAs), with their fine-grained programmability, offer a promising alternative. FPGAs can support model-specific architectures and leverage optimizations such as low-bitwidth quantization \cite{chen2023m4bram} and customized numerical formats \cite{wu2023msd}. 




\begin{figure}[h]
\centering
\includegraphics[width=0.9\linewidth]{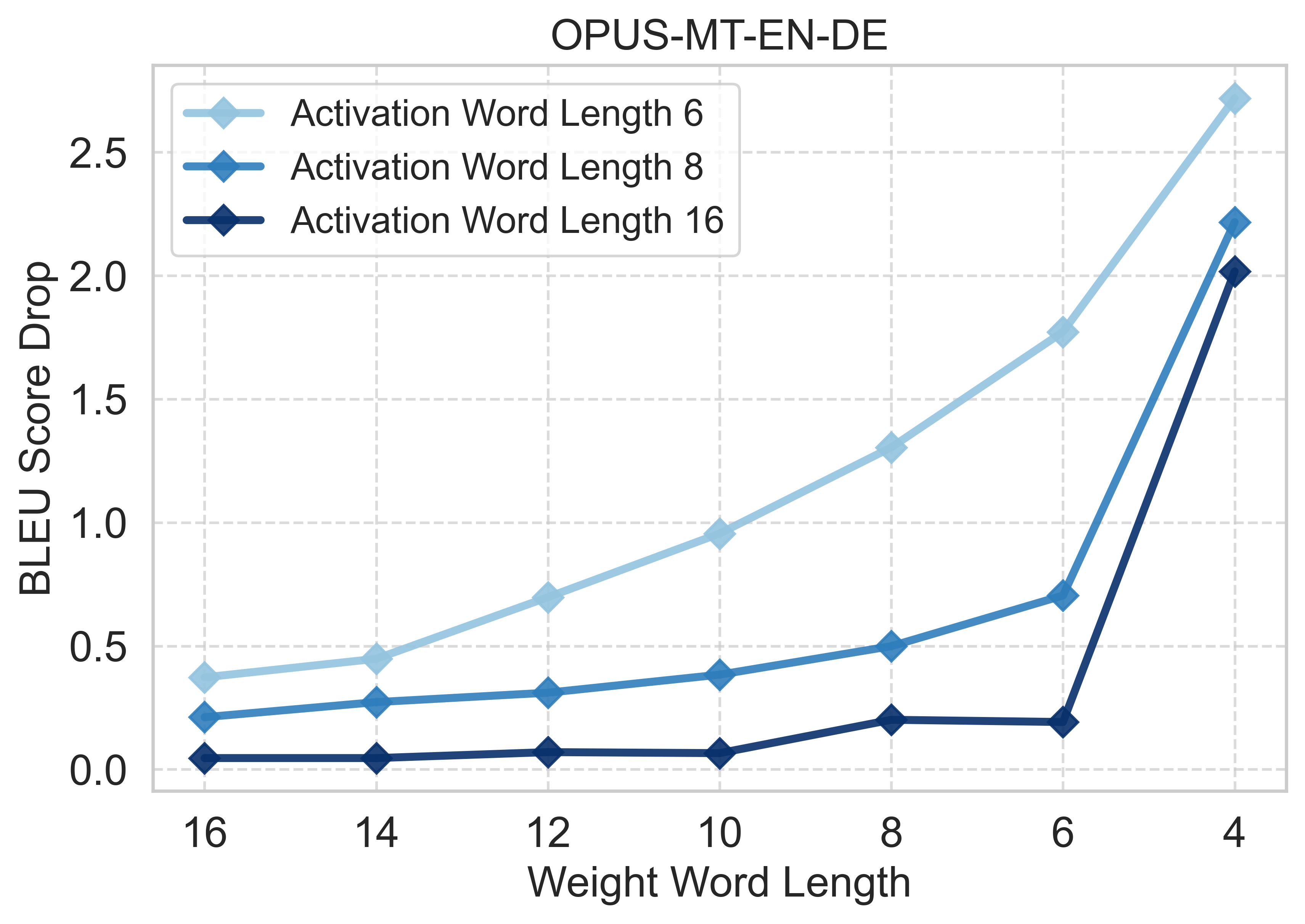}
\vspace{-10pt}
\caption{Post-training quantization results for an OPUS-MT model \cite{tiedemann2023democratizing} evaluated on a language translation dataset. The results demonstrate the reduction in the \texttt{BLEU Score} (a metric for translation accuracy) as model precision decreases, with the IEEE Single Precision (\texttt{FP32}) baseline serving as a reference. At the extreme quantization of \texttt{W4A8}, the \texttt{BLEU Score} is reduced by 2.22 (5.37\% compared to the \texttt{FP32} baseline).}
\label{img:quantization_only}
\vspace{-10pt}
\end{figure}

In LLM FPGA acceleration, post-training compression techniques have gained significant traction due to their ability to reduce computational and memory demands without requiring access to the original training data. However, many studies have shown that achieving effective LLM compression through quantization alone remains challenging, particularly when pushing model precision below \texttt{W8A8} (weights and activations in 8-bit fixed-point). State-of-the-art quantization methods with \texttt{W4A8} usually suffer from an average downstream task accuracy drop around 5\% compared to FP32 baseline \cite{shao2024omniquantomnidirectionallycalibratedquantization,liu2024qllmaccurateefficientlowbitwidth}. In contrast, Singular Value Decomposition (SVD) low-rank tensor decomposition as a post-training compression technique has been less explored in the context of LLMs, despite its substantial potential. The decomposition involves approximating the weight matrices in LLMs with low-rank matrices, reducing the model size and computational demands. In addition, SVD decomposition can be integrated with quantization schemes to enhance the efficiency of model compression.

\begin{figure*}[t]
\centering
\includegraphics[width=0.9\linewidth]{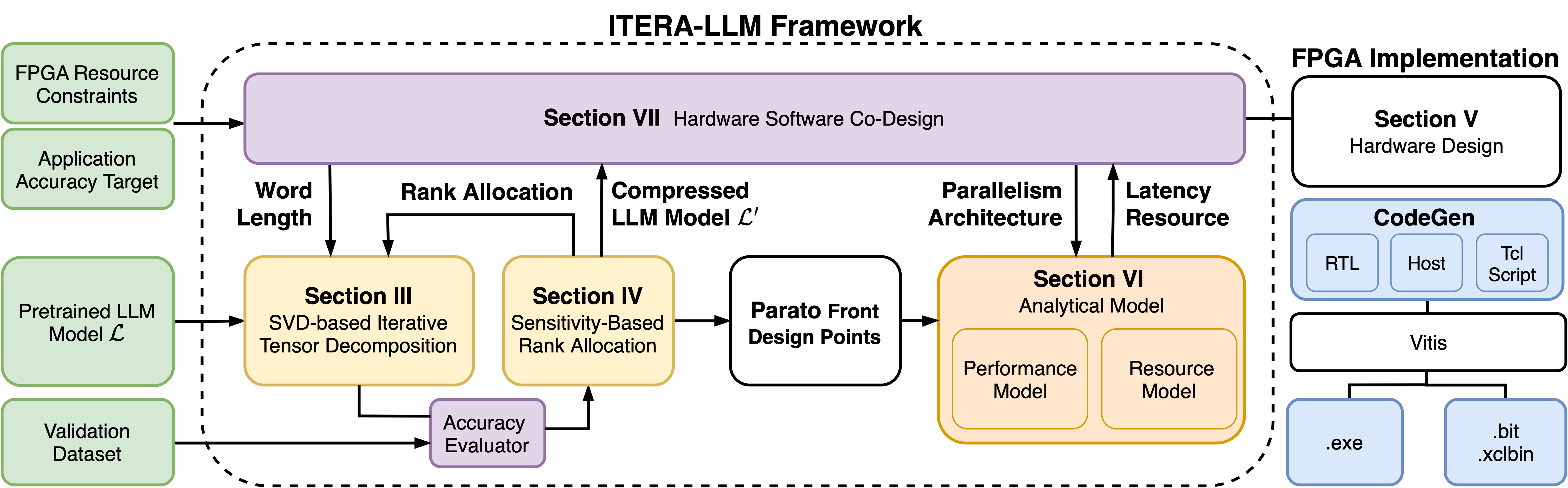}
\vspace{-5pt}
\caption{The overview of proposed \textit{ITERA-LLM} framework. This work focuses on the topic of post-training compression and FPGA accelerator co-design for producing Pareto-optimal design points on the accuracy-latency frontier. 
}
\vspace{-10pt}
\label{img:framework}
\end{figure*}

Furthermore, existing work often designs compression algorithms and hardware accelerators in isolation. This disconnect can lead to suboptimal solutions, as compression algorithms are predominantly designed to minimize the number of operations and parameters, without considering hardware-specific constraints. In this work, we propose \textit{ITERA-LLM}, a software-hardware co-design framework that integrates sub-8-bit quantization with SVD-based iterative tensor decomposition. Specifically, the work focuses on optimizing matrix multiplication (MatMul) operations in the linear layers, as these operations dominate the overall computational cost of LLMs \cite{ma2024efficientarbitraryprecisionacceleration}. The work assumes that these weight matrices are large and stored off-chip. The main contributions of this paper are as follows:

\begin{itemize}
    \item We introduce a novel LLM compression algorithm based on quantization, low-rank tensor decomposition, and sensitivity-based rank allocation. Our approach achieves up to \textbf{4.9\% improvement in model accuracy} at \texttt{W4A8} compared to existing quantization-only methods at a comparable compression ratio. Furthermore, for a similar model accuracy, our method \textbf{reduces the total number of fixed-point operations by 12.5\%} at \texttt{W6A8} compared to the quantization-only approach.

    \item We introduce a hardware-agnostic layer-wise optimization approach that maintains consistent bit-width across all layers, achieving model optimization through tuning the decomposition rank $r$ of each layer.
    \item To support hardware-aware deployment, we develop analytical models to estimate the performance and resource utilization of our compressed LLMs on FPGA platforms. These models support automated Design Space Exploration (DSE), enabling comprehensive evaluation of optimal solutions that balance accuracy and latency.

    \item We evaluate our \textit{ITERA-LLM} framework using OPUS-MT\cite{tiedemann2023democratizing} models. Under the resource constraints of ZCU111, our experiments show linear layer latency reductions ranging from \textbf{12.1\%} (0.879$\times$) to \textbf{41.1\%} (0.589$\times$), compared to the quantization-only MatMul baseline with comparable model accuracy.
\end{itemize}

\section{Related Work}
\label{sec:section2}
As quantizing both the weights and activations of LLMs to sub-8-bit fixed-point often results in severe accuracy degradation, various approaches have been proposed to address this challenge.

One such approach is Quantization-Aware Training (QAT). Q8BERT \cite{Zafrir_2019} emulated integer quantization operations during the training phase and employed the Straight-Through Estimator (STE) to approximate gradients for non-differentiable quantization operations. This method achieved 8-bit integer quantization on the BERT model, with no accuracy loss compared to the 32-bit floating-point representation. Similarly, Q-BERT \cite{shen2020q} reduced precision to 4 bits with less than 1\% accuracy loss. While QAT methods demonstrate promising sub-8-bit results, they also introduce significant challenges, including the high computational cost of training LLMs and the need for access to extensive datasets.

An alternative approach is to stay with post-training quantization while enhancing its performance through layer-wise optimization. For example, mixed-precision quantization \cite{shen2020q, lee2024owq} can be employed, where different layers of the model are quantized at varying bit-widths to balance overall accuracy and compression. The per-layer bitwidth decisions can be informed by sensitivity analysis, such as evaluating the gradient of the loss function with respect to the weight parameters \cite{dong2019hawq}. While layer-wise optimization can capture variations in each layer effectively, it often requires bit-serial hardware \cite{chen2024bitmod} to support varying precision levels. Such bit-serial designs are more favorable for ASICs but less efficient for FPGAs, as dedicated DSP cores cannot be fully utilized \cite{umuroglu2019optimizing}.

Another promising direction is combining quantization with other orthogonal compression techniques, such as pruning \cite{tareen2024hihispmv, yu2024hass} and tensor decomposition \cite{fan2022adaptable, yu2021streamsvd}. For this line of work, in addition to algorithmic integration, FPGA accelerators could serve as a suitable platform for efficiently implementing these combined techniques in hardware. However, current research in this combined direction has either focused on convolutional neural networks \cite{yu2021streamsvd} or optimizations for LLMs using 16-bit precision \cite{zhuang2023charm}, leaving sub-8-bit LLMs unaddressed.

In this paper, we address this research gap by investigating the combination of sub-8-bit quantization with SVD approximation on LLMs. An overview of our proposed \textit{ITERA-LLM} framework is shown in Fig.~\ref{img:framework}, with detailed descriptions of each component provided in the following sections. Additionally, we highlight key differences with related work in Table~\ref{Prior_Comparison}.



\definecolor{ForestGreen}{rgb}{0.13, 0.55, 0.13}
\begin{table}[h]
    \centering
    \caption{Comparison with related work.}
    \label{Prior_Comparison}
    \vspace{-5pt}
    \footnotesize
    \begin{tabular}{lcccccc}
        \toprule
        \makecell[cc]{} & 
        \makecell[cc] {\cite{yu2021streamsvd}} & 
        \makecell[cc] {\cite{sun2022film}} & 
        \makecell[cc] {\cite{shen2020q}} &
        \makecell[cc] {\cite{zhuang2023charm}} &
        \makecell[cc] {\cite{Chen_2024}} &
        \makecell[cc] {\textit{ITERA-LLM}}\\
        \midrule
        \makecell[cc]{\textbf{LLMs}} & \textcolor{red}{$\times$} & \textcolor{red}{$\times$}& \textcolor{ForestGreen}{\checkmark} &
        \textcolor{ForestGreen}{\checkmark} &
        \textcolor{ForestGreen}{\checkmark} &
        \textcolor{ForestGreen}{\checkmark}\\
        \midrule
        \makecell[cc]{\textbf{Post-Training}}& \textcolor{ForestGreen}{\checkmark} & \textcolor{ForestGreen}{\checkmark} &
        \textcolor{red}{$\times$} &
        \textcolor{ForestGreen}{\checkmark} &
        \textcolor{ForestGreen}{\checkmark} &
        \textcolor{ForestGreen}{\checkmark}\\
        \arrayrulecolor{black!30}\midrule
        \makecell[cc]{\textbf{Sub-8-Bit}} & \textcolor{red}{$\times$} & \textcolor{ForestGreen}{\checkmark} &
        \textcolor{ForestGreen}{\checkmark} & \textcolor{red}{$\times$} &
        \textcolor{ForestGreen}{\checkmark} &
        \textcolor{ForestGreen}{\checkmark}\\
        \arrayrulecolor{black!30}\midrule
        \makecell[cc]{\textbf{Layer-wise}\\\textbf{Optimization}} &  \textcolor{ForestGreen}{\checkmark} & \textcolor{ForestGreen}{\checkmark} &
        \textcolor{ForestGreen}{\checkmark} & \textcolor{red}{$\times$} &
        \textcolor{red}{$\times$} &
        \textcolor{ForestGreen}{\checkmark}\\
        \midrule
        \makecell[cc]{\textbf{SVD}} &  \textcolor{ForestGreen}{\checkmark} & \textcolor{red}{$\times$}& \textcolor{red}{$\times$} &
        \textcolor{red}{$\times$} &
        \textcolor{red}{$\times$} &
        \textcolor{ForestGreen}{\checkmark}\\
        \arrayrulecolor{black}\bottomrule
    \end{tabular}

    \vspace{-10pt}
\end{table}


\section{SVD-based Iterative Tensor Decomposition}
\label{sec:section3}
\subsection{SVD-based Tensor Decomposition}
Assume a linear projection layer of an LLM that projects a hidden dimension of size \( K \) to size \( N \), with batch size \( M \). This layer can be expressed as:
\begin{equation}
    \mathbf{Y} = \mathbf{XW}, \quad \mathbf{X} \in \mathbb{R}^{M \times K},  \mathbf{W} \in \mathbb{R}^{K \times N}, \, \mathbf{Y} \in \mathbb{R}^{M \times N}
\end{equation}
where \( \mathbf{X} \) is the activation matrix, \( \mathbf{W} \) is the weight matrix, and \( \mathbf{Y} \) is the output. SVD can be used to decompose and approximate the weight matrix, producing an approximation of the matrix-matrix product. In this context, SVD decomposes a weight matrix \( \mathbf{W}\) into three matrices: \( \mathbf{U} \), \( \boldsymbol{\Sigma} \), and \( \mathbf{V}\), where \( \boldsymbol{\Sigma} \) is a diagonal matrix, with the diagonal values in \( \boldsymbol{\Sigma} \) being the singular values of \( \mathbf{W} \), and \( \mathbf{U}\) and \( \mathbf{V} \) are the corresponding left and right singular vector matrices, respectively \cite{10.5555/264989}. An approximation can be obtained by keeping the \( r \) largest singular values while the rest are truncated.
\begin{equation}
    \mathbf{W} \approx (\mathbf{U}_r \boldsymbol{\Sigma}_r^{\frac{1}{2}})(\boldsymbol{\Sigma}_r^{\frac{1}{2}} \mathbf{V}_r^\top) = \mathbf{W}_1\mathbf{W}_2
\end{equation}
where \( \mathbf{W}_1 \in \mathbb{R}^{K \times r}, \mathbf{W}_2 \in \mathbb{R}^{r \times N}\). To leverage the benefits of this low-rank decomposition in terms of computational efficiency and memory usage, the activation \( \mathbf{X} \) is directly multiplied by the decomposed matrices without reconstructing the original weight matrix \( \mathbf{W} \). Thus, the forward pass of the linear layer can be rewritten as:
\begin{equation}
\label{eq:svd_multiply}
    \mathbf{Y} \approx \mathbf{XW} = (\mathbf{X}\mathbf{W}_1)\mathbf{W}_2
\end{equation}

Usually, quantization is applied to the resulting \(\mathbf{W}_1\) and \(\mathbf{W}_2\) matrices. This transforms the original single matrix multiplication into two sequential matrix multiplications with the possibility to reduce the total number of operations and model parameters by selecting a suitable decomposition rank.

\subsection{SVD-based Iterative Tensor Decomposition Algorithm}

\begin{figure}[t]
\centering
\includegraphics[width=\linewidth]{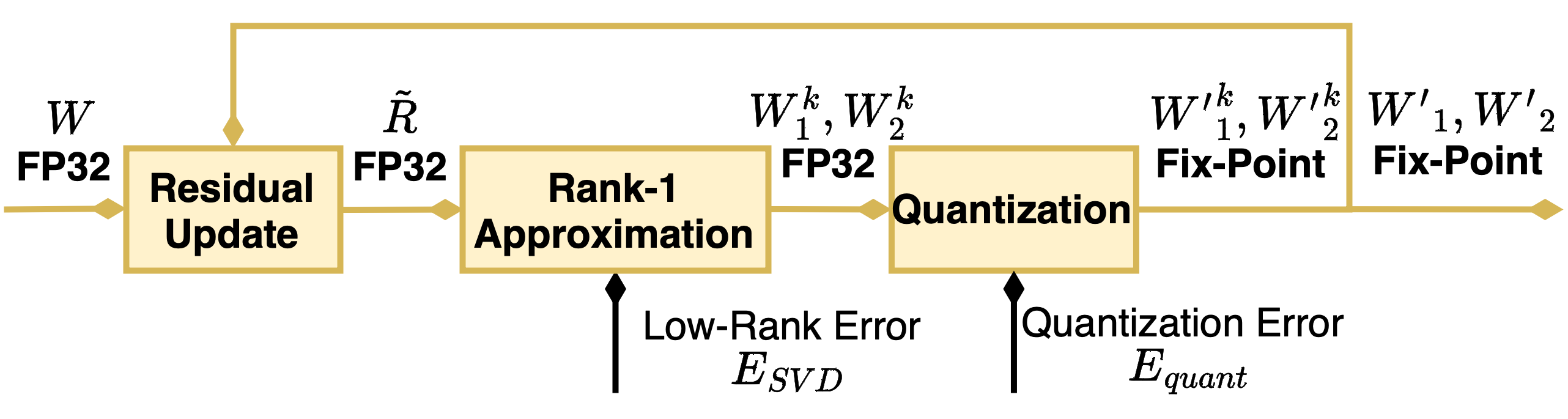}
\vspace{-15pt}
\caption{The proposed iterative tensor decomposition algorithm with SVD and quantization in a closed loop}
\label{img:iterative_quant_svd}
\end{figure}

Our SVD iterative decomposition transforms the traditional SVD decomposition into a refinement loop consisting of $r$ iterations.  In each iteration, the algorithm quantizes the singular vectors corresponding to the largest singular values. At the beginning, residual $\tilde{\mathbf{R}}$ is initialized as the weight $\mathbf{W}$. During subsequent iterations, the residual is updated by subtracting the product of the two quantized rank-1 singular vectors from the current residual. This iterative algorithm aims to minimize the Frobenius norm of the residual $\tilde{R}$ by producing two new quantized rank-1 singular vectors at each step. Outliers in the weight matrix with large magnitudes contribute disproportionately to the residual, forcing the algorithm to focus on these outliers for better approximation.

\begin{equation}
\| \tilde{\mathbf{R}}_r \|_F^2 = \| \mathbf{W} - \sum_{k=1}^r \mathbf{W'}_1^k \mathbf{W'}_2^k \|_F^2
\end{equation}
At the end of the loop, all quantized rank-1 singular vectors are augmented to constitute two quantized rank-$r$ matrices $\mathbf{W'}_1$, $\mathbf{W'}_2$. The details of the proposed iterative algorithm are illustrated in Fig.~\ref{img:iterative_quant_svd} and Algorithm \ref{alg:iterative_svd}.

\begin{algorithm}
\caption{SVD-based Iterative Tensor Decomposition}
\label{alg:iterative_svd}
\footnotesize
\begin{flushleft}
\textbf{Input:} $\mathbf{W}$, $\mathbf{r}$, \textbf{weight\_wl}\\
\Comment{\textcolor{magenta}{$\mathbf{W}$ is a FP32 weight matrix. $\mathbf{r}$ is the target rank for the decomposition, and \textbf{weight\_wl} is the word length of the quantization scheme.}} \\
\textbf{Output:} $\mathbf{W'}_1$, $\mathbf{W'}_2$\\
\Comment{\textcolor{magenta}{Output tensor decompositions are quantized with precision \textbf{weight\_wl}.}} \\
\end{flushleft}

\begin{algorithmic}[t]
\State Initialize $k \gets 1$
\State  $\mathbf{W}_1^1, \mathbf{W}_{2}^1 = \text{SVD}(\mathbf{W})_1$ \Comment{\textbf{Rank-1 Approximation}}
\State $\mathbf{W'}_1^1, \mathbf{W'}_2^1 = \text{Quant}(\mathbf{W}_1^1, \mathbf{W}_{2}^1,\text{weight\_wl})$ \Comment{\textbf{Quantization}}
\State $\tilde{\mathbf{R}} = \mathbf{W}-\mathbf{W'}_1^1\mathbf{W'}_2^1$ \Comment{\textbf{Residual Update}}
\State $\mathbf{W'}_1 = \left[\mathbf{W'}_1^1\right]$ and $\mathbf{W'}_2 = \left[\mathbf{W'}_2^1\right]$
\While{$k < r$}
    \State  $\mathbf{W}_{1}^k, \mathbf{W}_{2}^k = \text{SVD}(\tilde{\mathbf{R}})_1$ 
    \Comment{\textbf{Rank-1 Approximation} on Residual}
    \State $\mathbf{W'}_{1}^k, \mathbf{W'}_{2}^k = \text{Quant}(\mathbf{W}_{1}^k, \mathbf{W}_{2}^k,\text{weight\_wl})$ \Comment{\textbf{Quantization}}
    \State $\mathbf{W'}_1=\text{hstack}(\mathbf{W'}_1,\mathbf{W'}_{1}^k)$
    \State $\mathbf{W'}_2=\text{vstack}(\mathbf{W'}_2,\mathbf{W'}_{2}^k)$ \Comment{Augmentation}
    \State $\tilde{\mathbf{R}} = \tilde{\mathbf{R}}-\mathbf{W'}_{1}^k, \mathbf{W'}_{2}^k$ \Comment{\textbf{Residual Update}}
    \State $k \gets k + 1$
    
\EndWhile
\State return $\mathbf{W'}_1$, $\mathbf{W'}_2$
\end{algorithmic}
\end{algorithm}

\section{Sensitivity-Based Rank Allocation}
\label{sec:section4}
\begin{figure}[H]
\centering
\includegraphics[width=0.85\linewidth]{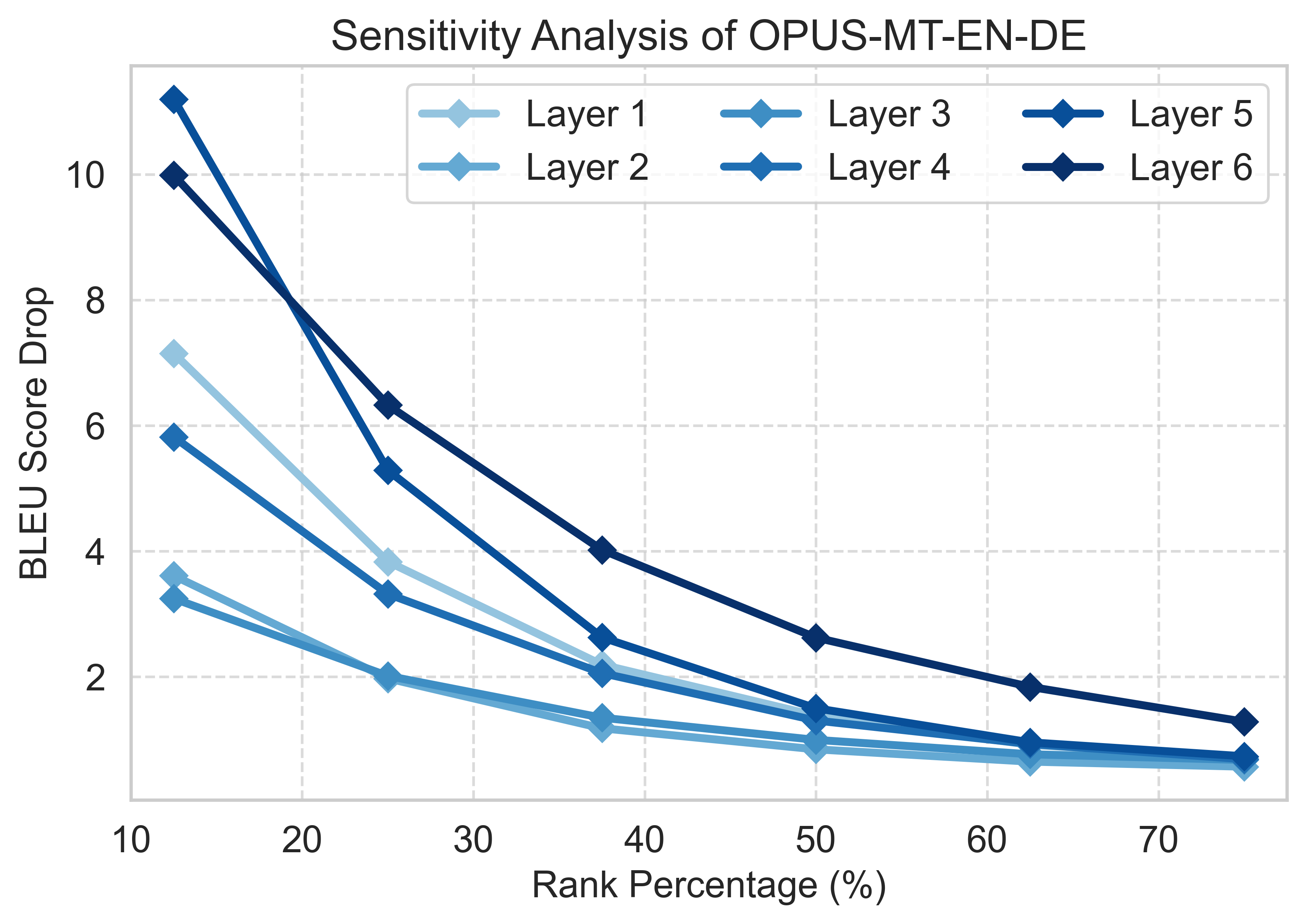}
\vspace{-5pt}
\caption{The sensitivity analysis measures the reduction in \texttt{BLEU Score} when varying the percentage of rank retained in each layer. Each layer's weight matrices are truncated to the specified rank percentage, temporarily replacing the original matrices in the model while keeping other layers unchanged.}
\label{img:sensitivity_analysis}
\vspace{-5pt}
\end{figure}

Different layers in LLMs exhibit varying degrees of sensitivity to rank truncation, as shown in Fig.~\ref{img:sensitivity_analysis}. While some layers can tolerate significant rank reduction, others are more sensitive and require higher ranks to preserve the overall model functionality. To address this variability, we introduce a novel Sensitivity-based Rank Allocation (SRA) algorithm that tunes the approximation of each layer of an LLM by adjusting the rank in its decomposition. The goal is to assign a rank for each layer so that the model achieves the best performance possible under a specific model compression ratio.

\subsection{Problem Formulation}

Let \( L \) represent the number of layers in the LLM. Denote the rank allocated to the \( i \)-th layer as \( r_i \), where \( r_i \in \mathbb{N} \). Our objective is to find the rank allocation \( [r_1, r_2, \dots, r_L] \) under the constraint that the total rank is equal to a given rank budget \( R_{\text{total}}^* \), and the model inference accuracy \(A\) is maximized. The optimization problem can be formulated as:
\begin{equation}
    \max_{[ r_1, r_2, \dots, r_L]} A \quad \text{subject to} \quad \sum_{i=1}^{L} r_i = R_{\text{total}}^*
\end{equation}

\subsection{Sensitivity-Based Rank Allocation (SRA) Algorithm}
The SRA algorithm iteratively allocates ranks to the layers of the LLM model \(\mathcal{L}\) based on their varying sensitivities. The algorithm's workflow is outlined below.
\begin{enumerate}
    \item \textbf{Initial Setup:} Initialize the rank \( r_i \) for each layer with  \( r_i = \frac{R_{\text{total}}^*}{L} \), such that the total rank budget is split equally across all layers.
    \item \textbf{Objective Evaluation:} For the given rank allocation \([r_1, r_2, \dots, r_L] \), the model accuracy \( A \) is evaluated using a randomly sampled calibration set.
    \begin{equation}
        A\leftarrow \mathcal{L}[r_1, r_2, \dots, r_L]
    \end{equation}
    \item \textbf{Sensitivity Approximation:} The sensitivity $S_{r_i}$ is defined as the partial derivative of model accuracy $A$ to the rank $r_i$. The idea is that layers with higher sensitivity will be allocated more ranks in order to achieve a larger improvement in accuracy.
    \begin{equation}
        S_{r_i}=\frac{\partial A}{\partial r_i} 
    \end{equation}
    Due to the non-linear nature of LLM, the model accuracy is not directly differentiable with respect to the rank of each layer, so we approximate the sensitivity $S_{r_i}$ using the finite difference method as:
    \begin{eqnarray}
    \begin{gathered}
    A_{i}^{+}\leftarrow \mathcal{L}[r_1, \dots r_i + \delta, \dots, r_L]\\
    A_{i}^{-}\leftarrow \mathcal{L}[r_1, \dots r_i - \delta, \dots, r_L]\\
    S_{r_i} \approx \frac{A_i^{+} - A_i^{-}}{2\delta}\\
    \end{gathered}
    \vspace{-0.25in}
    \end{eqnarray}
    where \( \delta \) is a small integer perturbation value.
    \item \textbf{Rank Adjustment:} Rank adjustment is performed iteratively. In each iteration, the layers with the highest and lowest sensitivities are identified, and their ranks are increased and decreased by \( \delta \), respectively.
    \begin{equation}
    \small{
        r_i^{\text{new}} = r_i + \delta \quad \text{for} \quad i = \arg \max_{i} \left[ S_{r_1}, S_{r_2}, \dots, S_{r_L} \right]
        }
    \end{equation}
    \begin{equation}
    \small{
        r_j^{\text{new}} = r_j - \delta \quad
        \text{for} \quad j = \arg \min_{j} \left[ S_{r_1}, S_{r_2}, \dots, S_{r_L} \right]
    }
    \end{equation}

    To improve the convergence property of the optimization, \( \delta \) decays over time during the iterations. The decay strategy is as follows:
    \begin{equation}
        \delta_{n} = \text{round}\left[\frac{\delta_0}{(1 + \alpha n)}\right]
    \end{equation}
    where \( \delta_0 \) is the initial perturbation value, \( n \) is the current iteration number, and \( \alpha \) is a small constant that controls the rate of decay. This decaying \( \delta \) ensures that the gradient approximation has a finer granularity as the algorithm approaches an optimal solution.
    \item \textbf{Termination:} The algorithm terminates when the \texttt{BLEU Score} converges to a maximum value or after a predetermined number of iterations.
\end{enumerate}

\section{Hardware Design}
\label{sec:section5}

\subsection{Tiling and Dataflow of a baseline MatMul engine}
\label{sec:section6.A}
\begin{figure}
     \vspace{-5pt}
     \centering
     \includegraphics[width=1\linewidth]{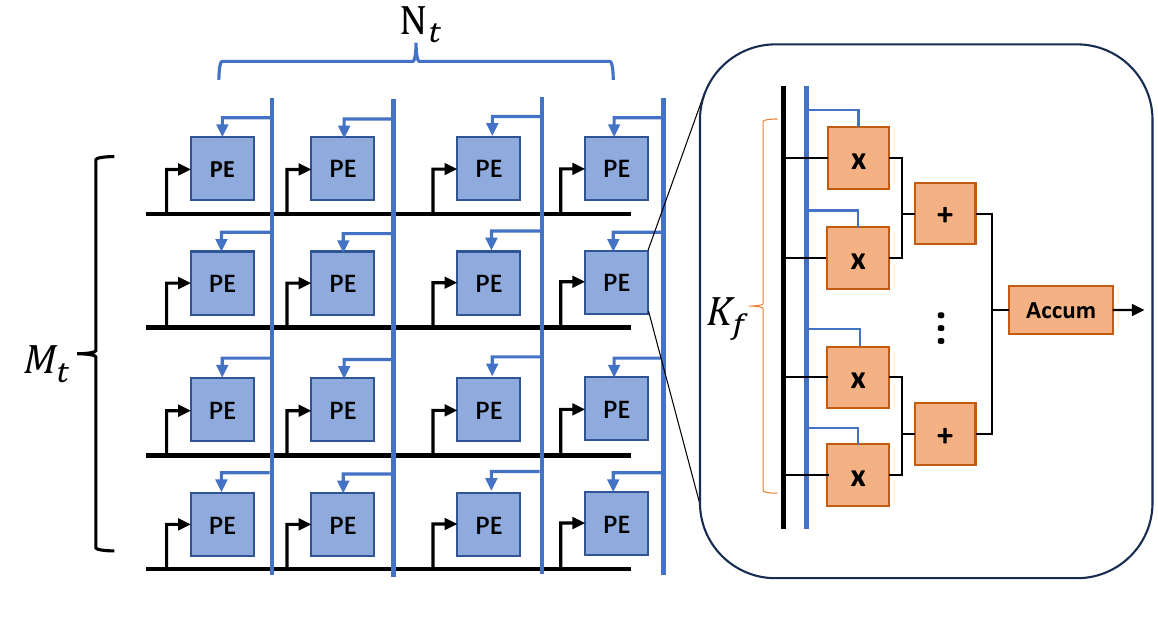}
     \vspace{-30pt}
     \caption{Dataflow and parallelism scheme for a target dense matrix-matrix multiplication layer.}
     \label{fig:matmul_dataflow}
\end{figure}

Fig.~\ref{fig:matmul_dataflow} and Listing~\ref{code:Data flow} describe the overall dataflow and the parameterization adopted in our basic MatMul engine for a baseline dense matrix-matrix multiplication $Y=XW$ operation of dimension $M \times K$ by $K \times N$. The basic MatMul engine applies parallelization on $M$, $K$, $N$ dimensions and tiling on $M$, and $N$ dimensions. The outer loop describes the tiling applied to the matrices. Given tiling factors of $M_t$ and $N_t$, the Left-Hand Side (LHS) and Right-Hand Side (RHS) matrices are broken down into tiles of $M_t \times K$ and $N_t \times K$ and are loaded from off-chip $M / M_t$ and $M / M_t \times N / N_t$ times respectively. At the \texttt{PE\_spatial\_loop}, we instantiate $M_t \times N_t$ processing elements in parallel. At the inner-most loop, each PE in the spatial array implements a Vector-Dot operation of $1 \times K$ and $K \times 1$ with a parallel factor of $K_f$. Overall the dataflow results in an output-stationary spatial array with $M/M_t \times N/N_t$ temporal loops. 

\begin{figure}[h]
\begin{lstlisting}[label=code:Data flow ,caption=Pseudocode of MM loop tiling and dataflow.]
tile_load_data_loop_M: for (int i.0=0; i.0<M/M_t; i.0++)
load_M_tile_from_offchip();
tile_load_data_loop_N: for (int j.0=0; j.0<N/N_t; j.0++)
    load_N_tile_from_offchip()
    PE_spatial_loop_M_t: for (int i.1=0; i.1<M_t; i.1++)
    PE_spatial_loop_N_t: for (int j.1=0; j.1<N_t; j.1++)
        PE_loop: for (int i.2=0; i.2<K/K_f; i.2++)
            parallel_dot_product();
\end{lstlisting}
\vspace{-0.25in}
\end{figure}

\subsection{MatMul with SVD scheduling}
As described in Equation~\ref{eq:svd_multiply}, since the original single MatMul layer requires two consecutive multiplications under the SVD approach, we propose two scheduling methods for the intermediate multiplication result using the MatMul engine described in Section~\ref{sec:section6.A}. In order to maximize performance, for both scheduling methods, we maintain intermediate results on-chip according to the corresponding tiling scheme.

\begin{figure}
     \centering
     \includegraphics[width=1\linewidth]{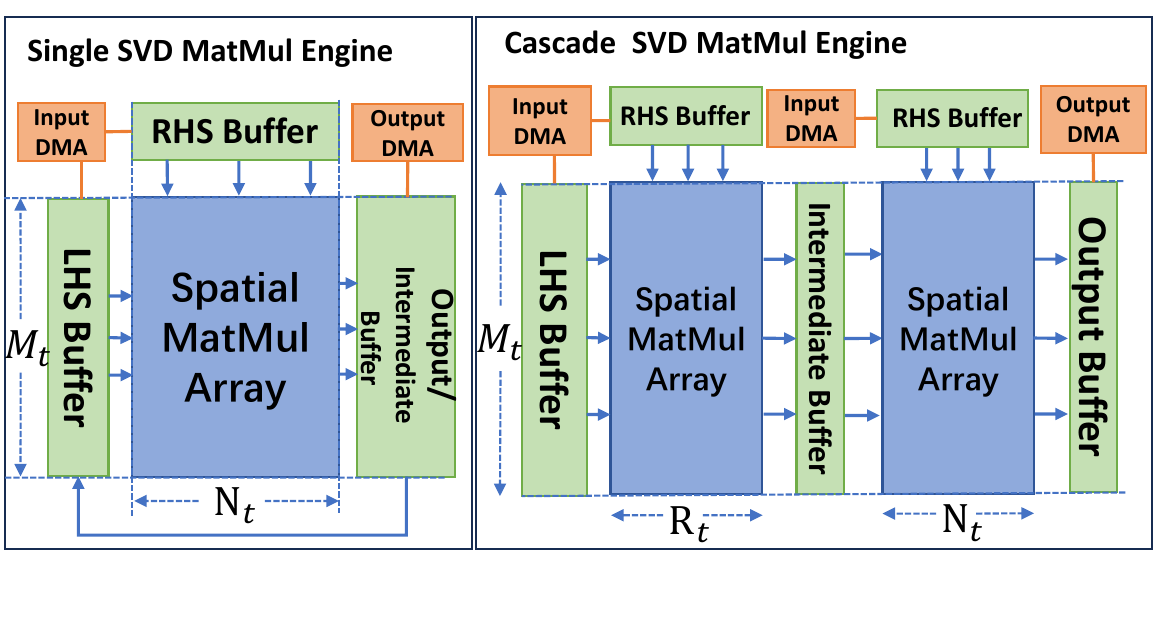}
     \vspace{-30pt}
     \caption{Single SVD MatMul engine architecture (left) vs. Cascade SVD MatMul engine architecture (right). The Spatial MatMul Array has the same architecture and parameterization as described in figure~\ref{fig:matmul_dataflow}. A connection to input/output DMA indicates where the accelerator communicates with off-chip memory. Blue arrows indicate on-chip communication.}
     \label{fig:single_vs_cascade}
     \vspace{-0.25in}
\end{figure}

\textbf{Single SVD MatMul Engine}: Fig.~\ref{fig:single_vs_cascade} (left) describes the overall architecture when mapping SVD workload to a single MatMul engine with a spatial tile size of $M_t \times N_t$. The single engine is reused temporally for the multiplication of $XW_1$ and $(XW_1)W_2$. The $N_t$ tiling factor is shared between the first part ($XW_1$) and the second part ($(XW_1)W_2$) of the multiplication, parallelizing both the $R$-dimension, with respect to the number of ranks, and the $N$-dimension, with respect to the output dimensionality. Since the tiling is applied to the outer loops ($M, R, N$ dimensions), this imposes a constraint that the entire tile of output from the first $XW_1$ multiplication with dimension $M_t \times R$ needs to be buffered on-chip for the subsequent multiplication which accumulates over the $R$-axis. 

\textbf{Cascade SVD MatMul Engine}: Fig.~\ref{fig:single_vs_cascade} (right) describes the architecture of a spatially unrolled cascade of MatMul kernels which implement the $XW_1$ and $(XW_1)W_2$ multiplication in parallel. By instantiating separate engines for the two multiplications we can assign separate tiling factors $R_t$ and $N_t$ for the $R$-dimension in the $K \times R$ ($W_1$) and the $N$-dimension in the $R \times N$ ($W_2$). We impose a constraint that both the preceding and succeeding MatMul engine must have the same $M_t$ tiling factor to match the dimensions without having to introduce any additional buffering. Similarly to the single MatMul engine, we need to buffer the entire $M_t \times R$ tile of intermediate results on-chip. 

\section{Analytical Modelling}
\label{sec:section_modelling}
In this section, we explain the analytical modelling framework adopted for fast prototyping and design space exploration, with the objective of identifying the hardware architecture and configuration that would lead to the lowest latency given the available resources. We adopt a bottom-up approach, modelling the performance and resource usage from a single PE to a spatial tile consisting of PEs. For simplicity, the discussion is based on a single dense matrix multiplication of $M \times K$ with $K \times N$, but it applies to both the Single SVD and Cascade SVD MatMul Engines for decoupled weight matrix. Each tile is configurable with parameters $M_t$, $N_t$ and $K_f$.

\subsection{Performance Modelling}
We adopt a rate and workload based approach in estimating the MatMul engine performance, where we model the input and output rates of each port of a hardware module expressed in the number of words per cycle and the workload each port produces or consumes expressed in the number of words. In the following discussion, we denote $r^i,r^o$ as the input and output rates, respectively, and $w^i,w^o$ as the input and output workloads, respectively. We denote with subscript $r^i_{LHS},r^i_{RHS}$, where input ports correspond to the workload from the Left-Hand-Side (LHS) matrix and the Right-Hand-Side (RHS) matrix, respectively.

\textbf{PE}: As described in Listing~\ref{code:Data flow}, each PE functions as a vector-matrix engine that performs multiply and accumulate operations along the $K$-dimension with a parallel input factor of $K_f$. As the multiply and accumulate units are fully pipelined, each PE has a rate of input and output of:
\begin{eqnarray}
\vspace{-0.30in}
\small{
\begin{gathered}
\label{eq:rate_pe}
r^i_{LHS} = \frac{K}{\ceil[\big]{\frac{K}{K_f}} \times N} \\
r^i_{RHS} = K_f \\
r^o = \frac{1}{\ceil[\big]{\frac{K}{K_f}}} \\
\end{gathered}
}
\vspace{-0.30in}
\end{eqnarray}


\textbf{Matrix Multiply Tile}: Building from the above rate models, and assuming $M_t$ and $N_t$ are tiling factors for the LHS and RHS matrix, respectively, the rate models for a MatMul tile can be obtained as:
\begin{eqnarray}
\vspace{-0.30in}
\small{
\begin{gathered}
\label{eq:rate_tile}
r^i_{LHS} = M_t \times \frac{K}{\ceil[\big]{\frac{K}{K_f}} \times N} \\
r^i_{RHS} = N_t \times K_f \\
r^o = M_t \times N_t \times \frac{1}{\ceil[\big]{\frac{K}{K_f}}} \\
\end{gathered}
}
\vspace{-0.30in}
\end{eqnarray}


Here, we define the workload for each port as the total number of words transferred in or out of the corresponding port. For a single tile, this can be obtained from Listing~\ref{code:Data flow} as:
\begin{eqnarray}
\small{
\begin{gathered}
\label{eq:workload}
w^i_{LHS} = M \times K \\
w^i_{RHS} = \frac{M}{M_t} \times K \times N \\
w^o = M \times N \\
\end{gathered}
}
\vspace{-0.25in}
\end{eqnarray}


Finally, the latency of a MatMul tile can be obtained from the maximum number of cycles producing/consuming the corresponding workload:
\begin{equation}
    \label{eq:latency}
    \small{
        latency = \max\left[\frac{w^i_{LHS}}{r^i_{LHS}}, \frac{w^i_{RHS}}{r^i_{RHS}}, \frac{w^o}{r^o}\right]
    }
\end{equation}

\subsection{Resource Modelling}

\textbf{DSP}: The DSP utilization model captures the number of multipliers instantiated in parallel in each PE. Where ($f_{packing}$)\cite{chen2023m4bram} defines the number of multiplications packed into a single DSP, the total DSP utilization is given by:
\begin{eqnarray}
\small{
\begin{gathered}
\label{eq:dsp_model}
DSP_{PE} = \ceil[\big]{\frac{K_f}{f_{packing}}} \\
DSP_{tile} = M_t \times N_t \times DSP_{PE} \\
\end{gathered}
}
\vspace{-0.25in}
\end{eqnarray}

\textbf{BRAM}: On-chip buffer is allocated to each tile according to the tiling factors assigned to the LHS and RHS matrices. To enable parallel processing on the $K$-axis, and since the dual-ported BRAMs are configured as FIFOs between inputs from off-chip and DSPs, each DSP is assigned a single BRAM for parallel access. We denote here $\textbf{bram18(}Buff_{depth}, \text{bitwidth}\textbf{)}$ as a modelling function for the number of BRAM18K units instantiated through synthesis for a buffer array with depth $Buff_{depth}$ with corresponding bitwidth. The total on-chip memory requirement in terms of BRAM18K is given by:
\begin{equation}
    \label{eq:buffer_depth}
    \small{
    Buff_{depth} = \ceil[\big]{\frac{K}{K_f}}
    }
    \vspace{-0.2in}
\end{equation}
\begin{eqnarray}
\small{
\begin{gathered}
\label{eq:bram_modelling}
BRAM_{PE} = \ceil[\big]{\frac{K_f}{f_{packing}}} \times \text{bram18}(Buff_{depth}, \text{bitwidth}) \\
BRAM_{LHS} = M_t \times BRAM_{PE} \\
BRAM_{RHS} = N_t \times BRAM_{PE} \\
\end{gathered}
}
\vspace{-0.25in}
\end{eqnarray}


\textbf{Off-chip Bandwidth}: The off-chip bandwidth requirement for a given tile is the average bits/cycle required for the MatMul tile to run at full throughput. This is obtained from the total workload transferred between on-chip and off-chip divided by the total latency running the workload:
\begin{equation}
\small{
    Bandwidth = \frac{w^i_{LHS} + w^i_{RHS} + w^o}{latency}
    }
\end{equation}


\section{Hardware-Software Co-Design}
Designing hardware accelerators and compression algorithms separately can lead to suboptimal solutions. To address this challenge in a structured manner, the hardware software co-design framework shown in Fig.~\ref{img:framework} has been developed to identify a set of design points that achieve better accuracy-latency trade-offs under hardware resource constraints. 

\begin{itemize}
    \item \textbf{Model Compression and Pareto Analysis}:  
    The framework iterates through different bit-widths to compress the given LLM model using our iterative SVD tensor decomposition and sensitivity-based rank allocation algorithm. This process determines the optimal quantization precision of the entire model and decomposition rank for each layer. Two Pareto fronts are identified:  
    \begin{itemize}
        \item \textbf{Accuracy vs. Number of Operations}  
        \item \textbf{Accuracy vs. Model Compression Ratio}  
    \end{itemize}

    \item \textbf{Hardware-Aware Design Space Pruning}:  
    Using the constraints of the target FPGA platform (e.g., DSPs, BRAMs, off-chip bandwidth), the framework prunes the design space by eliminating configurations that exceed available hardware resources. 

    \item \textbf{Hardware-Aware Performance Exploration}:  
    For each design point on the Model Pareto front, performance and resource models are instantiated to find the hardware configuration with the lowest latency by sampling the constrained hardware design space. Based on whether the platform is compute-bound or memory-bound, the framework generates a set of compressed models \(\mathcal{L}'\) and their corresponding hardware accelerators \(H\), which jointly provide an optimized accuracy-latency trade-off for a target FPGA platform.

\end{itemize}
This framework generates model compression tailored to the unique characteristics of the compute platform. By integrating algorithmic flexibility with platform-specific optimizations, our approach achieves superior accuracy-latency trade-offs as demonstrated in Section~\ref{sec:section8} Evaluation.




\section{Evaluation}
\label{sec:section8}
\subsection{Experimental Setup}
The FPGA device selected for implementation and performance evaluation is the ZCU111, with Vitis HLS 2023.2 used to synthesize the designs. The clock frequency of our accelerators is configured to 200 MHz. The proposed framework is evaluated using a family of transformer-based neural machine translation (NMT) models called OPUS-MT \cite{tiedemann2023democratizing}. We evaluate our framework on two specific source-to-target language pairs: English-to-German (EN-DE) and French-to-English (FR-EN) using the \texttt{WMT2019} dataset. The accuracy of machine translation is evaluated using BiLingual Evaluation Understudy (\texttt{BLEU Score}).

\subsection{Quantization and SVD Baselines}
For evaluation, we compared our work against a state-of-the-art quantization-only baseline. Specifically, we compare our approach to an OPUS-MT model compressed using a post-training quantization scheme \cite{Zafrir_2019}, which is then deployed on a highly optimized systolic array accelerator as implemented in \cite{Chen_2024}. The notation \texttt{WXAY} represents a weight word length of $X$ and an activation word length of $Y$. We use the same quantization scheme for all the methods we evaluate. To highlight the performance improvements of our iterative decomposition, we also present an SVD tensor decomposition baseline. The \texttt{FP32} weight matrices are first decomposed using SVD, followed by quantization of the produced matrix to the target word length. For a fair comparison, quantization is applied vector-wise in the produced matrix to align with the implementation in our SVD-based iterative tensor decomposition. The SVD baseline adopts a uniform decomposition rank across all linear layers in the model and does not use our proposed SRA algorithm.


\subsection{Evaluation on Model Compression}
This work focuses on boosting \textbf{sub-8-bit} post-training compression. We evaluate our approach against baseline approaches in terms of \textbf{accuracy}, \textbf{compression ratio}, and \textbf{number of operations (NOps)}. The methods compared include:
\begin{itemize}
    \item Quantization baseline
    \item SVD tensor decomposition
    \item \textbf{Ours: SVD iterative tensor decomposition}
    \item \textbf{Ours: SVD iterative tensor decomposition with SRA}
\end{itemize}

\begin{figure}[t]
\centering
\includegraphics[width=\linewidth]{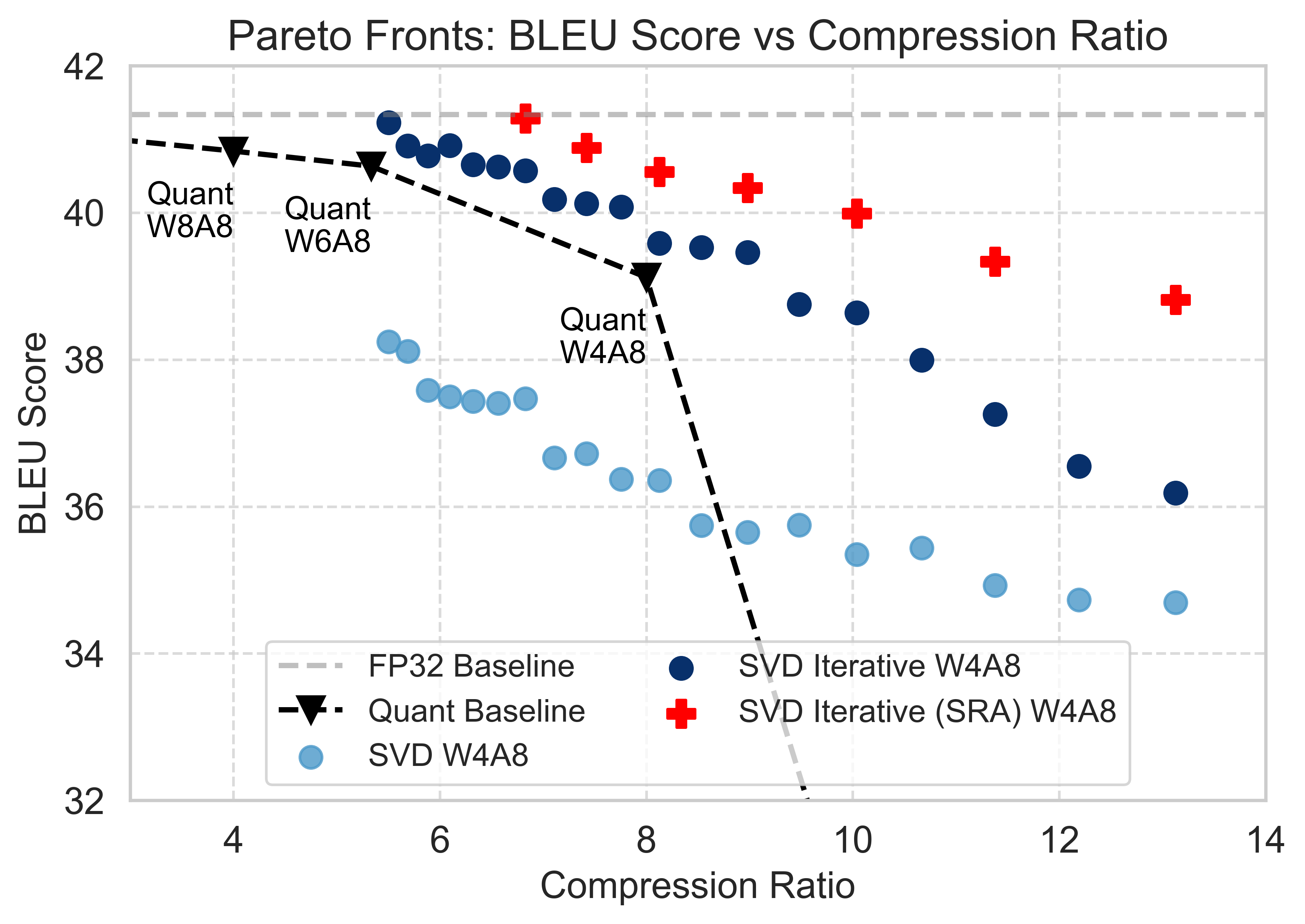}
\vspace{-20pt}
\caption{Pareto fronts of \texttt{BLEU score} (accuracy) versus model compression ratio obtained from design space exploration. The Pareto fronts correspond to the design points of SVD Iterative (SRA) \texttt{W4A8}, highlighted in red with a cross symbol.}
\label{img:svd_quant_compression_ratio}
\vspace{-0.25in}
\end{figure}

The compression ratio is normalized relative to the \texttt{FP32} model size. For instance, a compression ratio of 4 corresponds to 8-bit fixed-point quantization. Therefore, compression ratios greater than 4 fall within our region of interest. As shown in Fig.~\ref{img:svd_quant_compression_ratio}, the SVD tensor decomposition approach underperforms the quantization baseline in the region of interest, only surpassing it at very low compression ratios. In contrast, our iterative SVD tensor decomposition outperforms both the SVD tensor decomposition and quantization baseline across the entire spectrum of compression ratios. This approach compensates for quantization errors through the residual mechanism in the iterative refinement loop. The iterative process progressively mitigates the cumulative error from both low-rank approximation and quantization, effectively improving model accuracy by refining the weight representation at each iteration. 

However, the accuracy gain from SVD iterative tensor decomposition \texttt{W4A8} diminishes as the compression ratio increases. The higher compression ratio corresponds to a smaller rank of the SVD decomposition, which results in fewer refinement iterations. Finally, we compare the SVD iterative tensor decomposition with the SVD iterative tensor decomposition with SRA. The SRA approach provides a more substantial accuracy gain, particularly at low compression ratios. This is due to the model’s increased sensitivity to rank variation at lower compression ratios. By optimizing the rank allocation, the SRA approach efficiently captures the most critical ranks within the available rank budget, thereby preserving the model's inference capacity. 

\begin{figure}[t]
\centering
\includegraphics[width=\linewidth]{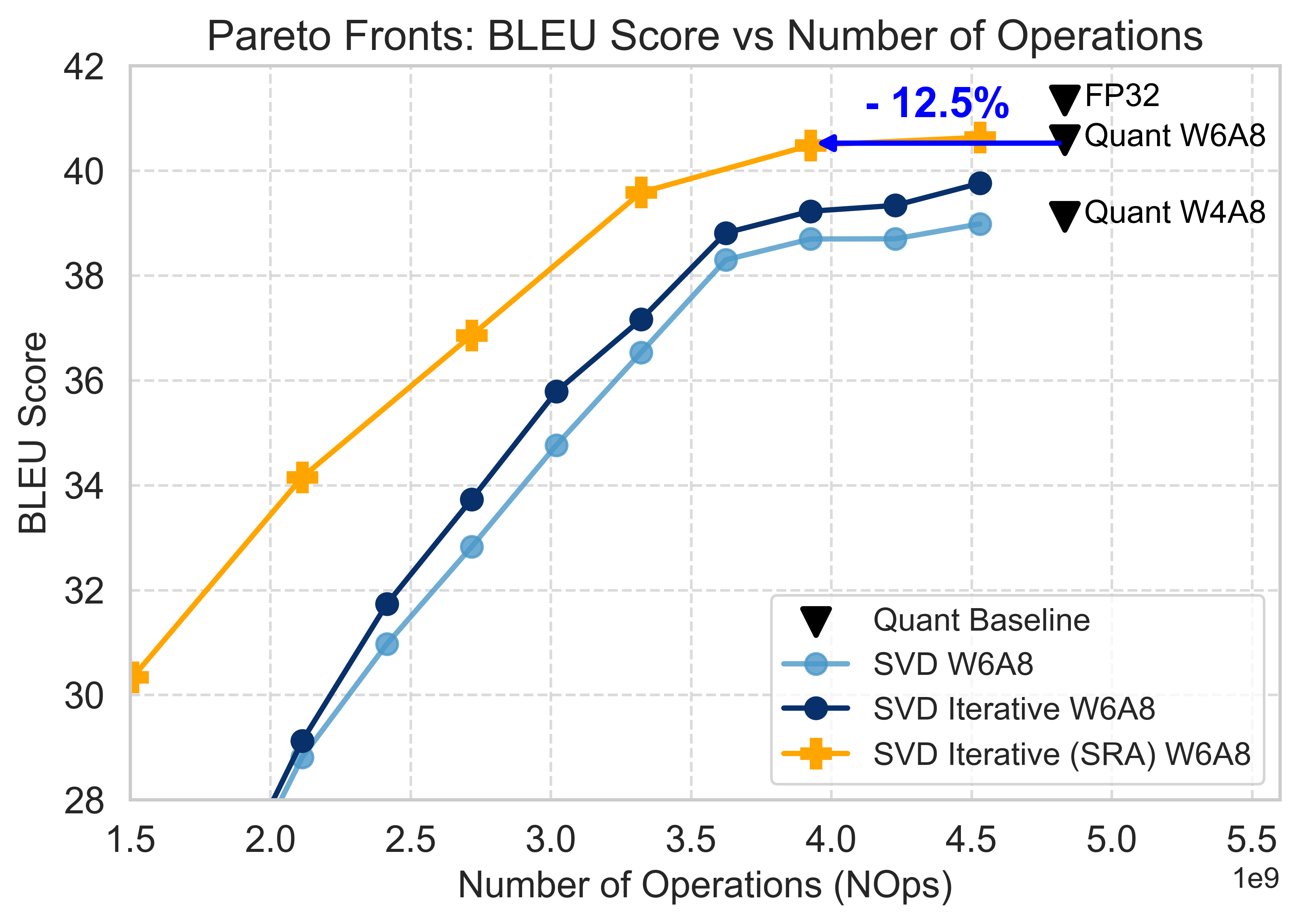}
\vspace{-20pt}
\caption{Pareto fronts of \texttt{BLEU score} (accuracy) versus number of operations obtained from design space exploration. The Pareto fronts correspond to the design points of SVD Iterative (SRA) \texttt{W6A8}, highlighted in orange with a cross symbol.}
\label{img:svd_quant_nops}
\vspace{-10pt}
\end{figure}

A similar argument applies to the Pareto fronts for the \texttt{BLEU Score} versus the number of operations. As shown in Fig.~\ref{img:svd_quant_nops}, the Pareto fronts are shaped by SVD iterative tensor decomposition with SRA \texttt{W6A8}. For a similar model accuracy, our method reduces the total number of fixed-point operations by 12.5\% at \texttt{W6A8} compared to the quantization baseline.



\subsection{Single MatMul Engine vs. Cascade MatMul Engine}
Using the hardware modelling framework, we compared the Pareto performance of Single MatMul Engine and Cascade MatMul Engine against a MatMul baseline without applying SVD under the resource constraints of ZCU111 and different off-chip bandwidth requirements. We select a workload of $M \times K \times N = 512 \times 512 \times 512$, where the weight matrix $K \times N$ has a full-rank of 512. This is a dominant workload for the $Q, K, V$ (Query, Key, Value) layers in the attention heads for the OPUS-MT model. 

\begin{figure}[t]
\centering
\includegraphics[width=\linewidth]{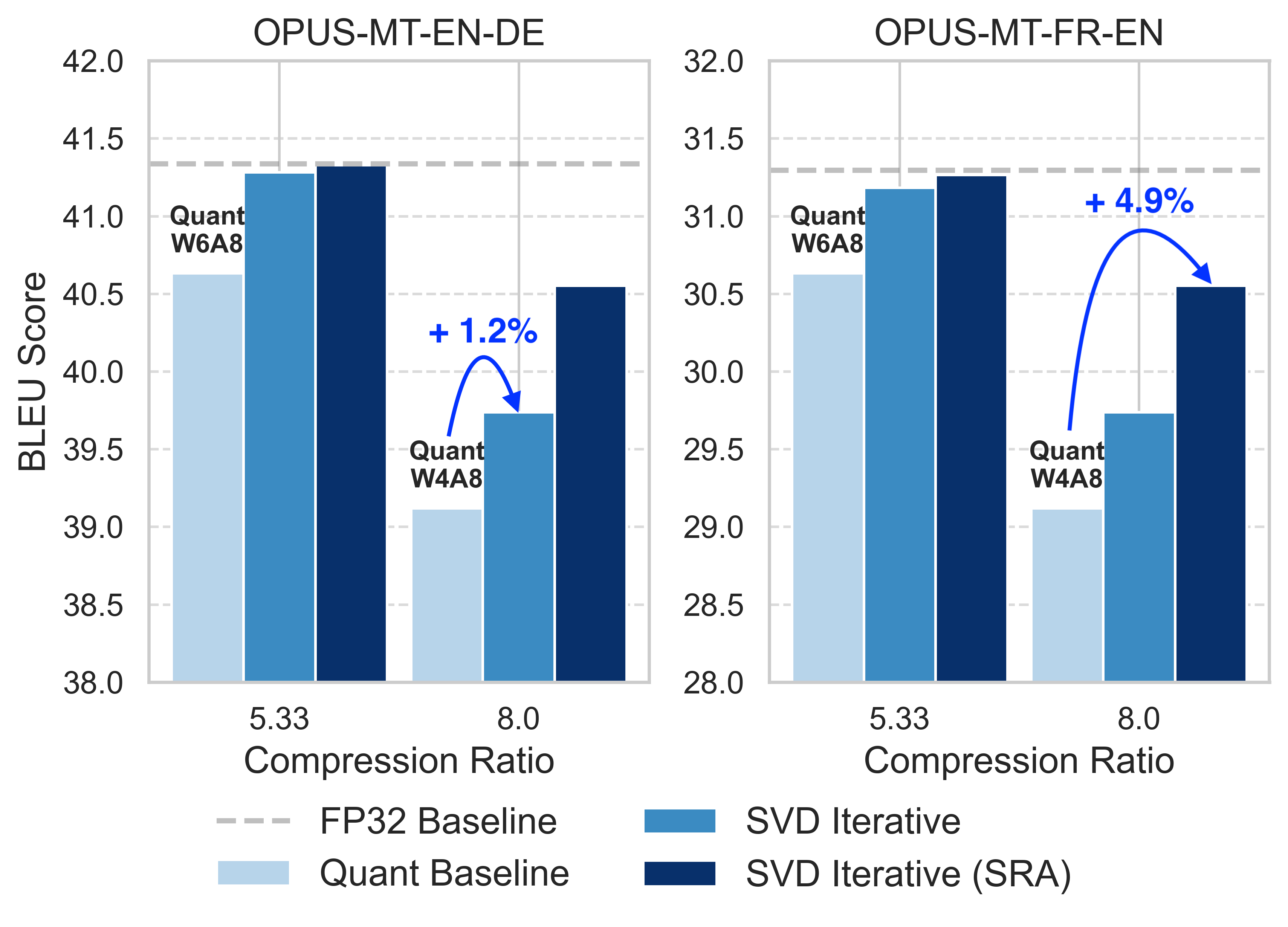}
\vspace{-20pt}
\caption{Bar plot of \texttt{BLEU score} versus compression ratio for OPUS-MT models with different source-to-target language pairs, demonstrating the generality of our approach. At compression ratio 8, our iterative SVD tensor decomposition \texttt{W4A8} improves accuracy by 1.2\% over quantization-only \texttt{W4A8}, while SRA further boosts accuracy by up to 4.9\% compared to quantization.}
\label{img:svd_compaison}
\vspace{-0.25in}
\end{figure}

\begin{figure}
    \centering
    \includegraphics[width=\linewidth]{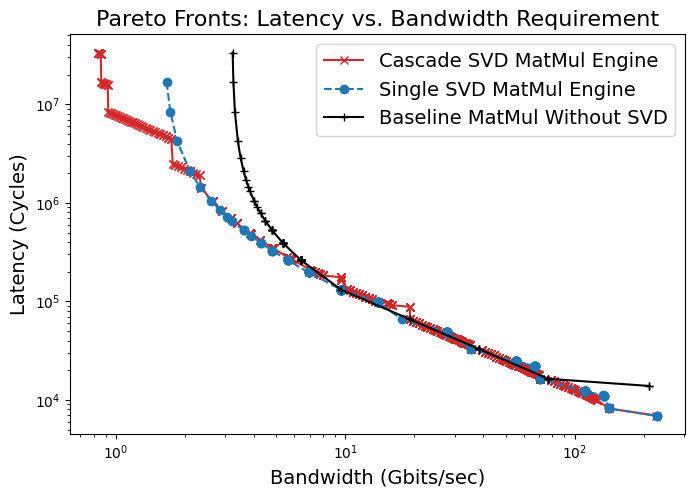}
    \vspace{-20pt}
    \caption{Pareto fronts of different modes of MatMul engines' latencies and corresponding bandwidth requirement to run at full throughput, evaluated with $M \times K \times N=512 \times 512 \times 512$ (W4A8). SingleTile and CascadeTile were evaluated at rank=128, under a resource constraint of DSP=4272, BRAM18K=1080 (ZCU111).}
    \label{fig:latency_vs_bandwidth}
    \vspace{-10pt}
\end{figure}

Fig.~\ref{fig:latency_vs_bandwidth} illustrates the target operational bandwidth range for SVD MatMul engines. In the bandwidth-limited region on the left-hand side of the spectrum, the SVD MatMul engines achieve comparable latency to the baseline MatMul engine with reduced off-chip bandwidth requirements. This is attributed to the use of lower-rank decomposed weight matrices, which reduce off-chip data transfers. As we move toward the right-hand side, the design space transitions from being memory-bound to compute-bound. In this region, SVD MatMul engines achieve lower latency under the same bandwidth, owing to the reduced number of operations required for low-rank matrix computations. Furthermore, we notice that the Cascade engine populates a finer design space in between the Pareto points of Single SVD engine due to having finer-grained parameterization between the two consecutive SVD workloads. We therefore include both Single and Cascade MatMul engines to lead a finer grained DSE process.

\begin{figure}
    \centering
    \includegraphics[width=0.9\linewidth]{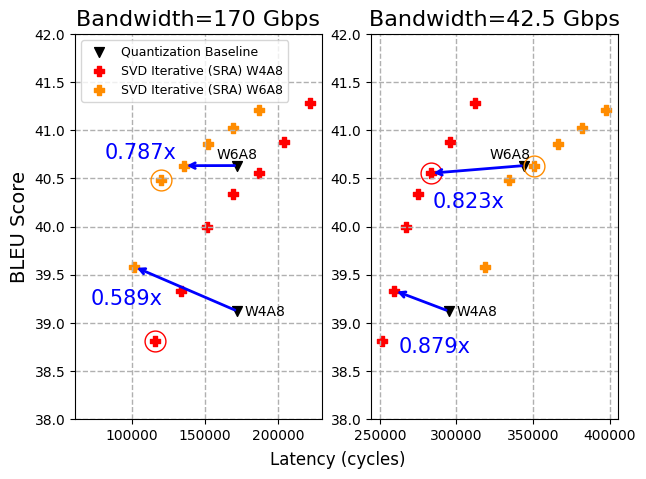}
    \vspace{-10pt}
    \caption{Trade-offs between \texttt{BLEU score} and performance (latency) across compression methods mapped onto MatMul kernels, evaluated with batch size 512 under resource constraints of ZCU111 under two different off-chip bandwidth constraints. \textbf{Left}: original bandwidth of ZCU111, \textbf{Right}: A quarter of the original bandwidth to simulate an extreme bandwidth-limited situation. Latency comparison of design points with comparable \texttt{BLEU score} marked with blue arrows.}
    \label{fig:bleu_vs_latency}
    \vspace{-0.2in}
\end{figure}




\begin{figure}
    \centering
    \includegraphics[width=\linewidth]{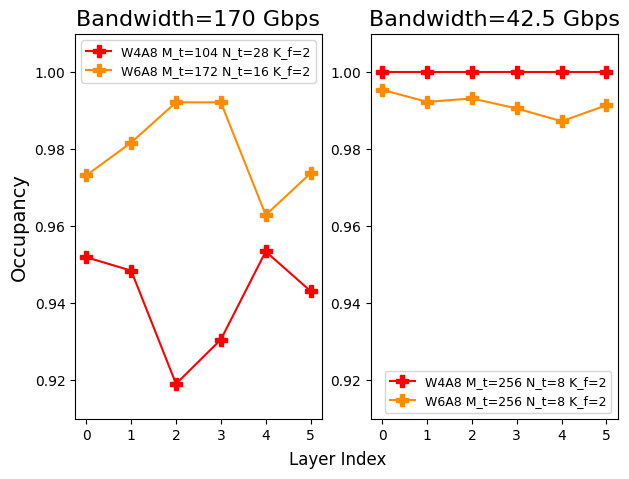}
    \vspace{-20pt}
    \caption{Layer-wise MatMul Tile occupancy for selected design points (marked with circle) from Fig.~\ref{fig:bleu_vs_latency}.}
    \label{fig:layer_wise_occupancy}
    \vspace{-10pt}
\end{figure}

\newpage
\subsection{Mapping compression methods onto MatMul engines}
In this section, we evaluate the performance of Pareto design points identified in Fig.~\ref{img:svd_quant_compression_ratio}, in terms of compression ratio, and Fig.~\ref{img:svd_quant_nops}, in terms of number of operations. For the SVD iterative approach with SRA, each layer is assigned a unique rank. We explore a range of SVD MatMul engine configurations (both Single and Cascade) across the entire hardware design space, and measure per-layer latency for each configuration. The configuration that yields the lowest total latency is selected as the optimal accelerator design point. For the quantization baseline, we similarly evaluate a set of MatMul engine configurations with varying parameterizations for each fixed-point quantization scheme, and report the lowest total latency achieved for each scheme.

Fig.~\ref{fig:bleu_vs_latency} (left) demonstrates when the SVD MatMul engine operates within the bandwidth requirement (compute-bounded). In this case, the SVD iterative (SRA) \texttt{W6A8} outperforms \texttt{W4A8} in terms of \texttt{BLEU Score} and latency as a higher bit-width enables a lower decomposition rank per layer and hence fewer operations. In Fig.~\ref{fig:bleu_vs_latency} (right), we simulate a bandwidth-limited situation. In this case, the Pareto front corresponding to SVD iterative SRA \texttt{W4A8} outperforms all other compression methods as the bandwidth-limited platform favors a compression scheme with a higher compression ratio. Furthermore, we select four designs from Fig.~\ref{fig:bleu_vs_latency}, and provide a more detailed per-layer occupancy breakdown in Fig.~\ref{fig:layer_wise_occupancy}. For a given tile-size configuration, the occupancy variation between layers remains small ($<$5\%). This is because the achieved tile sizes on the target FPGAs are relatively small compared to the overall MatMul dimensions, minimizing the overhead introduced when padding is needed. We also observe that the bandwidth-limited scenario (Fig.~\ref{fig:layer_wise_occupancy}, right) tends to achieve higher occupancy compared to the compute-bounded scenario (Fig.~\ref{fig:layer_wise_occupancy}, left). This is because the hardware DSE framework selects smaller tile sizes that better align with the available memory bandwidth, making the side effect of padding less significant. In both compute-bounded and memory-bounded cases, design points from our SVD iterative (SRA) approach outperform the quantization baseline at comparable \texttt{BLEU Score}. By identifying the Pareto compression method while taking into account the hardware architecture and target platform resources, our framework generates a set of design points with improved accuracy-latency trade-offs in both compute-bound and memory-bound scenarios.




\section{Conclusion}
This paper proposes a model-accelerator co-design framework \textit{ITERA-LLM}. The framework considers simultaneously the compression of an LLM model through an iterative approximation scheme and the optimization of the hardware architecture taking into account the computational structure of the approximation scheme. As a result, \textit{ITERA-LLM} outperforms existing quantization-only work by producing design points that have a better accuracy-latency trade-off in both memory and compute-bound cases. 


\bibliographystyle{unsrt} 

\bibliography{reference} 

\end{document}